# Fabrication and installation of the Mu2e cryogenic distribution system

**M White, M Lamm, A Hocker, D Arnold, G Tatkowski, J Kilmer, V Poloubotko, T Tope, Y Huang, L Elementi, K Badgley, E Voirin, I Young, J Brandt, S Feher, C Hess, and D Markley**

Fermi National Accelerator Laboratory, PO Box 500, Batavia, IL 60510

Email: mjwhite@fnal.gov

**Abstract**. The muon-to-electron conversion (Mu2e) experiment at Fermilab will be used to search for the charged lepton flavor-violating conversion of muons to electrons in the field of an atomic nucleus. The Mu2e experiment is currently in the construction stage. The scope of this paper is the cryogenic distribution system and superconducting power leads for four superconducting solenoid magnets: Production Solenoid (PS), an Upstream and Downstream Transport Solenoids (TSu and TSd) and Detector Solenoid (DS). The design of the cryogenic distribution system and the fabrication of several sub-systems was reported previously. This paper reports on additional fabrication and installation progress that has been performed over the past two years. Lessons learned during fabrication and testing of the cryogenic distribution system components are described. In particular, the challenges and solutions implemented for aluminum welding are reported.

A description of the process used to qualify the welding procedure and welders for welding the aluminium stabilized NbTi superconducting power leads is provided. Additionally, the progress made with regards to installing the power leads into the cryogenic Feedboxes is covered.

## 1. Introduction

2. The liquid nitrogen ($LN_2$) storage dewar and helium refrigerators are located in the G-2 building and connected to the Distribution Box (DB) in the Mu2e building via a 150 m long outdoor transfer line running over the M4 beamline into the Mu2e building. The DB is connected to four Feedboxes via U-tube connections. The American Superconductor Corporation (ASC) and High Intensity Neutrino Source (HINS) solenoid power leads are installed in the Feedboxes and spliced to the superconducting bus that extends through the transfer line to each solenoid. The Feedboxes contain the valves and instruments to control the flow through each solenoid. The details of design of the cryogenic distribution system and the fabrication of several sub-systems was reported previously [1]. This paper reports on additional fabrication and installation progress that has been performed over the past two years. Lessons learned during fabrication and testing of the cryogenic distribution system components are described. In particular, the challenges and solutions implemented for aluminum welding are reported.

## 2. Cryogenic Distribution Box

Over the past two years the DB has been connected to the outdoor transfer line and 3,000 L liquid helium (LHe) Dewar as shown in figure 1. All necessary control valves have been fully installed and instrumentation has been connected. The plan is to cooldown the outdoor transfer line, DB, and 3,000 L LHe dewar in the fall of 2021 to commission these subsystems.

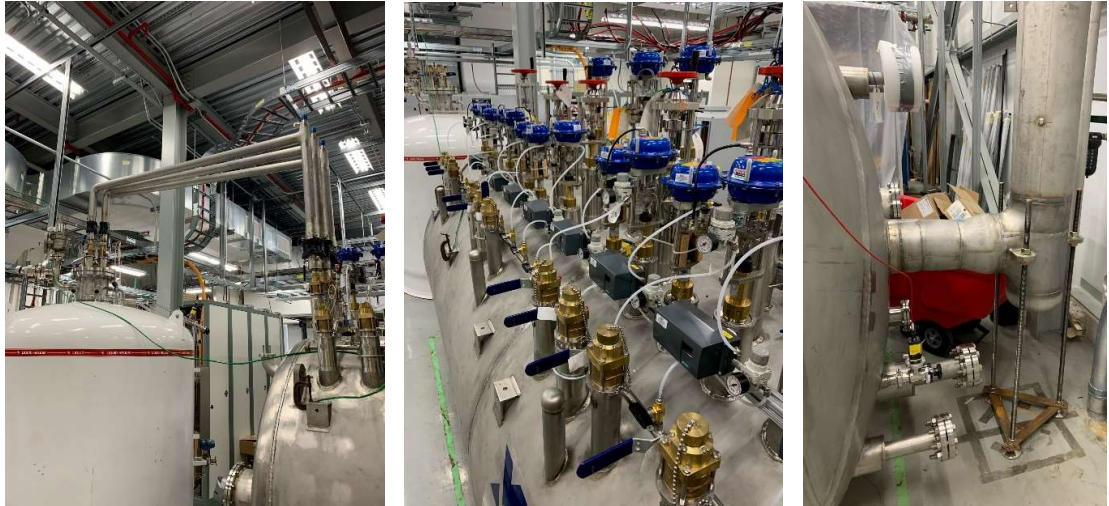

**Figure 1.** Photos showing the DB U-tube connections to the 3,000 L LHe dewar (left), the instrumentation and control valves installed on the DB (middle), and the DB connection to the outdoor transfer line (right).

### 3. Cryogenic Feedboxes

The design of the four cryogenic Feedboxes was previously reported [1]. Photos of the Feedboxes during the manufacturing process are shown in figure 2. The Fermilab specified maximum permitted integral helium leak rate is $1.3 \times 10^{-9}$ mbar*l*s$^{-1}$ for internal piping-to-vacuum as well as for air-to-vacuum as measured by a leak detector pumping on the vacuum jacket. The Feedbox manufacturer has decades of experience in helium leak checking and supplied specific detailed leak check procedures and records to Fermilab. All four Feedboxes successfully passed factory acceptance testing criteria

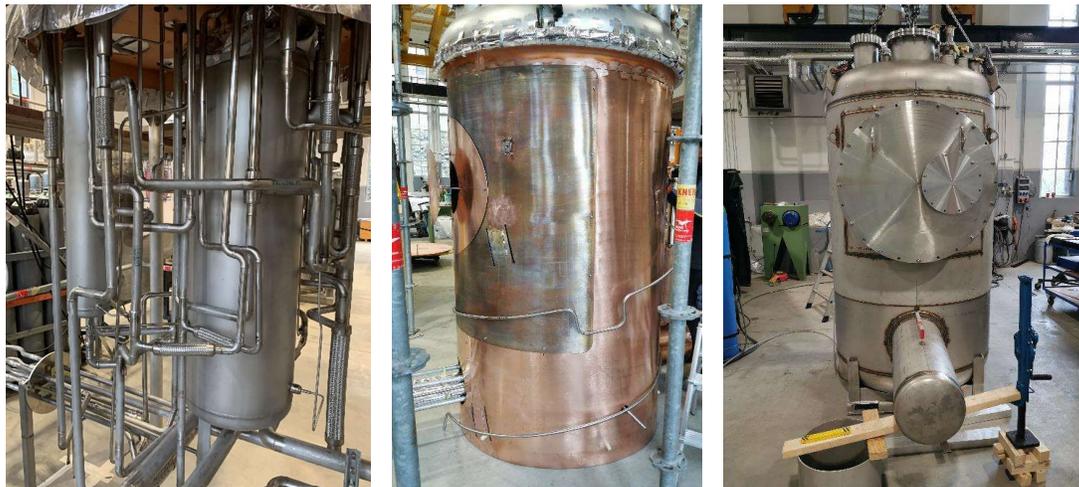

**Figure 2.** Photos showing the internal piping and vessels (left), thermal shield (middle), and vacuum jacket (right) during fabrication by the vendor.

Fermilab repeated the same factory acceptance testing helium leak check procedures and applied the same leak tightness requirements to the delivery acceptance tests upon delivery of the Feedboxes to Fermilab as shown in figure 3. No leak was detectable on the piping circuits when pressurized to their full design pressure with helium. The design pressure is 23.5 bar.a for the helium circuit and 11.4 bar.a for the LN$_2$ circuit.

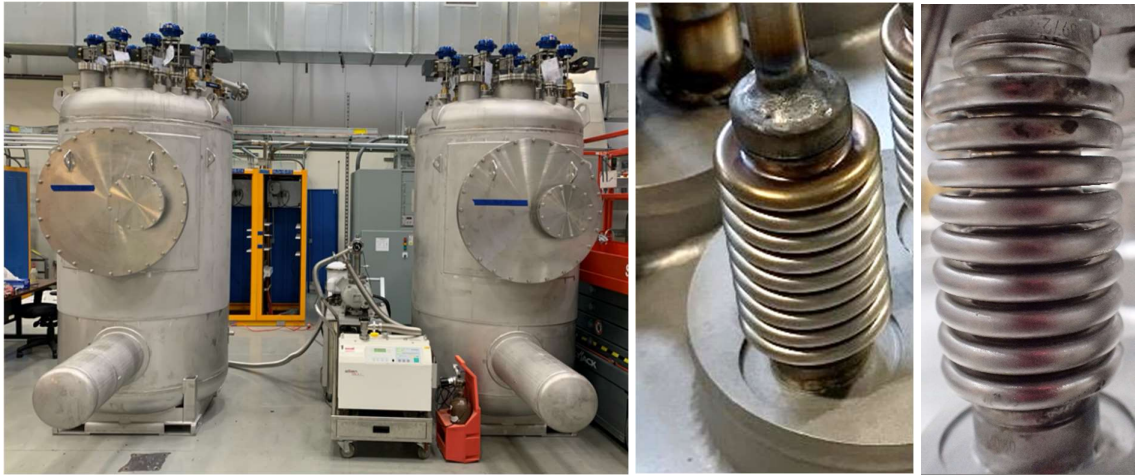

**Figure 3**. Photo showing the PS Feedbox (left) and DS Feedbox (right) during delivery acceptance testing performed upon the delivery of the Feedboxes to Fermilab in the Fall of 2020 (left). Photo showing one of the PS Feedbox liquid level port bellows during fabrication (middle) and after delivery to Fermilab (right)

Three of the four Feedbox insulating vacuum jackets similarly passed the $1.3 \times 10^{-9}$ mbar*l*s$^{-1}$ maximum helium leak rate requirement when bagged and sprayed with helium externally. A large leak on the PS Feedbox vacuum jacket liquid level ports was identified during preliminary leak checking. The manufacturer had applied pickling paste to remove the excessive heat tint on the bellows and the fillet weld to the inner helium tube. However, the pickling paste was not thoroughly neutralized, which is the probable cause of the black spots that had appeared by the time the Feedboxes were delivered to Fermilab. The bellows had two plies of 0.2 mm thickness, so the corrosion was able to generate a leak through both plies of the bellows. Fermilab expended significant effort to remove all black spots from various locations on the Feedboxes to prevent future corrosion issues. In addition, the two liquid level port bellows had to be replaced. Fermilab plans to specifically prohibit the use of pickling paste in future procurements of cryogenic equipment. The better solution is to control heat input, shielding gas, and other weld parameters to reduce the amount of heat tint generated.

The Feedboxes are designed for half of the insulating vacuum jacket to be removed to install the power leads and connect to the transfer line superconducting bus. Photos showing the removed sections being removed are shown in figure 4. The dust generated by the extensive grinding required is a hazard to personnel and adjacent experiment electronics racks. A large rolling plastic tent frame was constructed to contain the dust generated by grinding activities and a confined space blower was used to vent the dusty air outside. The high ventilation rate simplified personnel breathing protection during grinding and prevented dust from spreading.

Fermilab and several cryogenic equipment vendors with extensive and successful experience with stainless steel welding that were building Mu2e cryogenic system components had significant difficulties reliably generating aluminum piping welds that were both helium leak tight and compliant with piping code requirements for allowable internal weld protrusion. The manufacturer design originally included butt welded pipe elbows to connect supply and returns for testing at the ends of the transfer line stub, but helium leak tightness issues encountered on aluminum welds led to the design change to the lyra-shaped bends on the transfer line stub shown in figure 4 to minimize the number of required aluminum welds and maximize tube bending radii. Aluminum 6061-T6 is less ductile and tubing bends require more generous bending radii relative to typical austenitic stainless-steel materials used in cryogenics. The challenges with the aluminum welding on the Feedbox bimetallic joints, which must be kept less than 300 °C during welding, are described in figure 5.

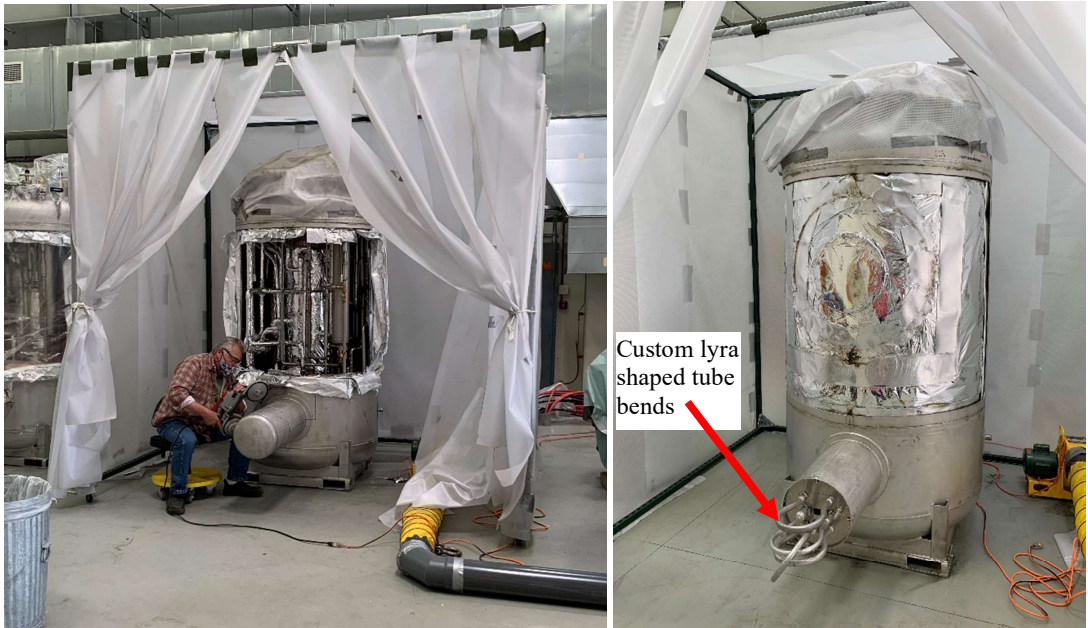

**Figure 4.** Photographs of the Feedbox transfer line stub (left) and just after the large removable section of the vacuum jacket was removed (right).

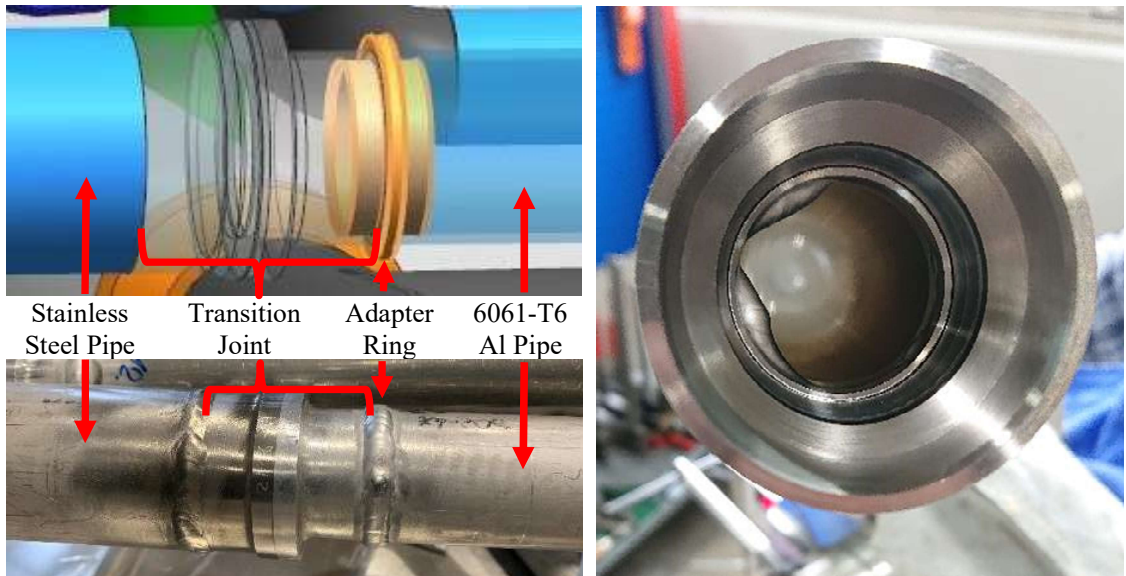

**Figure 5.** A machined aluminum adapter ring was utilized for the connection of the explosion bonded bimetallic transition joint to aluminum piping (left top and left bottom). The objective was to use small fillet welds on either side of the machined adapter rather than to directly butt weld to the transition joint to limit internal protrusion and heat input. In most cases the adapter ring completely melted into the weld metal. Photo of the worst internal weld protrusion on any of the 24 aluminum adapter rings (right).

## 4. Power leads and Superconducting Bus

The design and fabrication of the power lead subassemblies was previously reported [1]. The power leads were successfully test fit into the Feedboxes as shown in figure 6. The piping connections to the power leads within the Feedboxes are shown in figure 7.

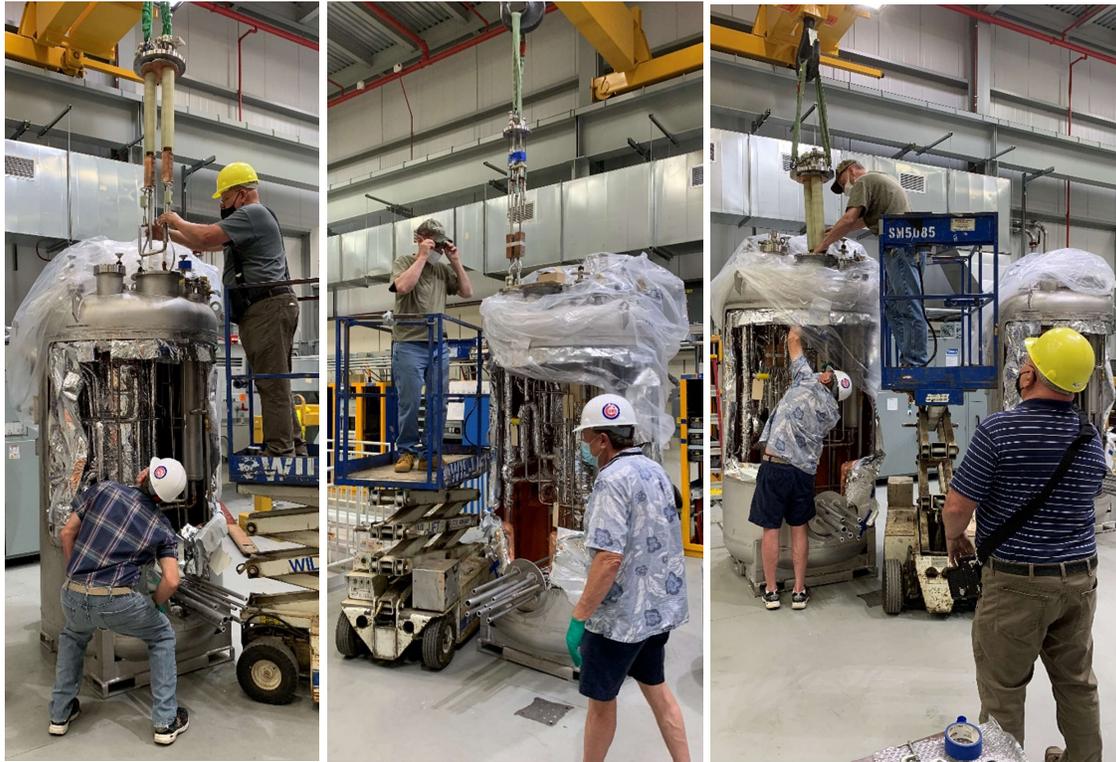

**Figure 6.** Successful test fit of the ASC leads into the PS Feedbox (left), the HINS leads into the TSd Feedbox (middle), and ASC leads into the TSd Feedbox (right).

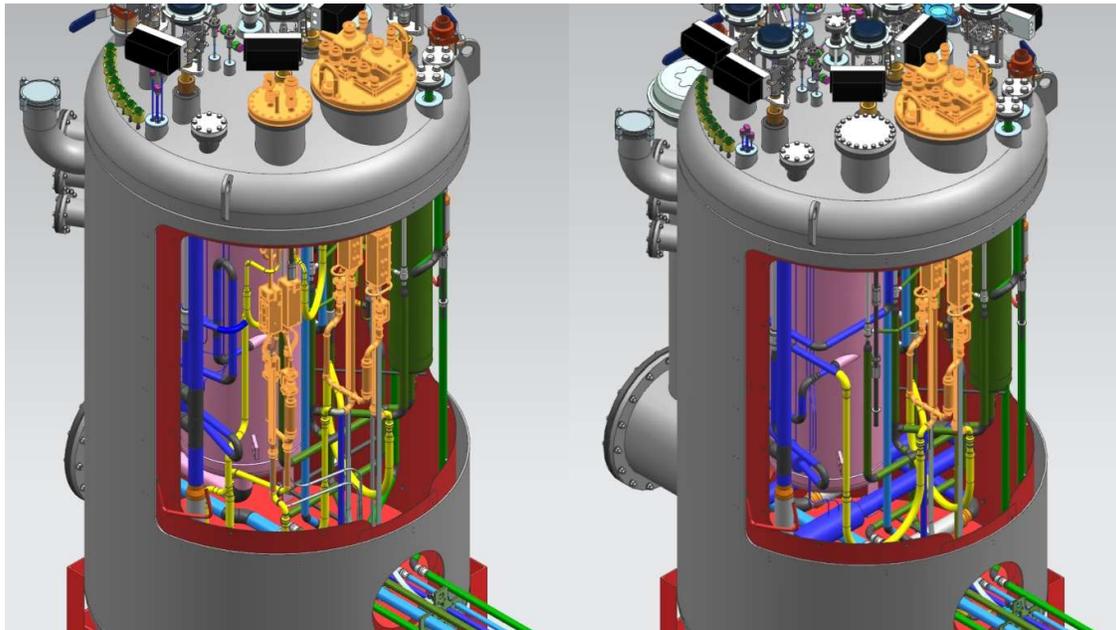

**Figure 7.** Picture of 3-D model showing the installation of the power leads (orange) and new piping (yellow) on the TSu/TSd Feedbox (left) and the PS/DS Feedbox (right). The warm end of lead has an $LN_2$ intercept and the cold end of the lead has a LHe intercept for both the ASC and HINS leads types. Large radius U-shaped flexhose bends are used to minimize forces applied to ceramic isolators.

A welding qualification campaign was conducted to ensure that Fermilab could reliably generate Gas Tungsten Arc Weld (GTAW) splice joint welds of superconducting bus with minimal electrical resistance. A set of 8 splices was cooled to LHe temperature, tested up to 10 kA, and measured to have resistances in the range of 0.2 to 0.7 n Ω. Minimizing the splice joint electrical resistance reduces the heat generated when the solenoids are powered and therefore reduces the number of thermal intercept clamp assemblies required. There is a splice joint between the power leads and the superconducting bus in the Feedbox, splice joints between transfer line sections, and final bus splice to the solenoid. In addition, the welder will be required to weld these splice joints in the flat, horizontal, vertical, and overhead welding position. Water-cooled copper blocks are required to prevent damage to the NbTi superconducting wire within the aluminum stabilizer as shown in figure 8.

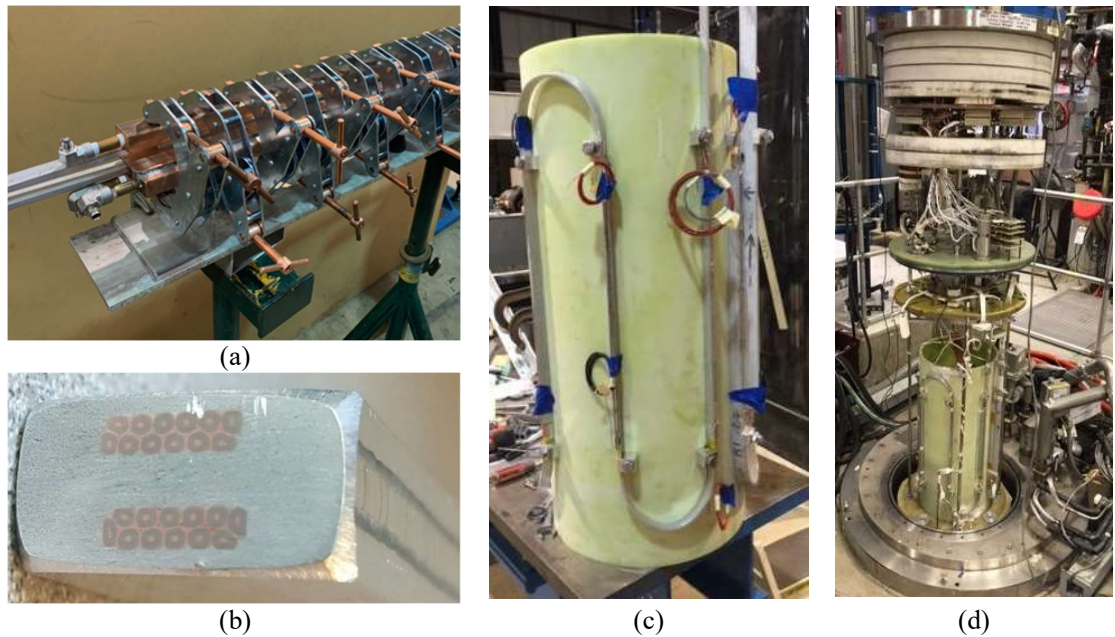

**Figure 8.** Photos of the superconducting bus cooling blocks and clamps (a), cross section of a practice PS horizontal fusion weld splice (b), the 8 qualification superconducting splices mounted on a G10 fixture (c), and the superconducting bus splices being tested in a cryostat at Fermilab (d).

**5. Solenoid Transfer Line**
The Solenoid Transfer Line (STL) aluminum piping welds were designed with knowledge of issues encountered on the TSu solenoid aluminum welds that are shown in figure 9. Similar problems were encountered with STL butt welds of aluminum pipe. The STL pipe joint design was modified into a socket-style coupling as shown in figure 10. The socket coupling design requires twice as many welds per joint location, but mitigates the risk of internal protrusion causing a pressure drop that interferes with the LHe thermosiphon cooling schemes used on the PS and DS solenoids. The transfer lines are installed with a slight downward slope between the Feedbox and PS and DS solenoids to help induce the desired thermosiphon flow rates. There are no flexible hoses in the STL, so the pipe elbows are the most highly thermally stressed locations in the STL internal piping. Custom elbows with added straight lengths were used to keep the weld joints away from the high thermal stress locations on the elbows.

Photos of the aluminum thermal shield and austenitic stainless steel vacuum jacket fabrication are shown in figure 11. A photo showing current progress on installing STL sections within the Mu2e lower detector hall is shown in figure 12.

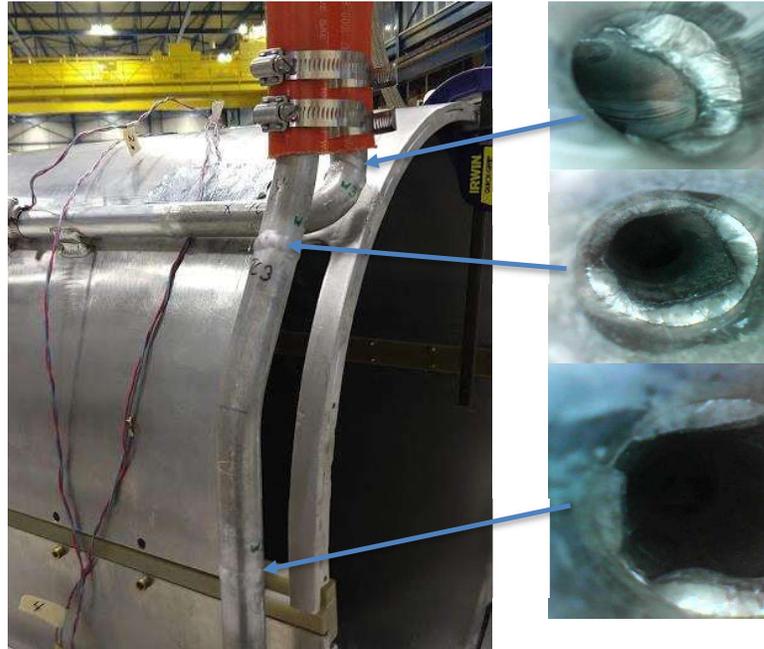

**Figure 9.** Examples of internal weld protrusion on the TSu Solenoid aluminum piping butt welds

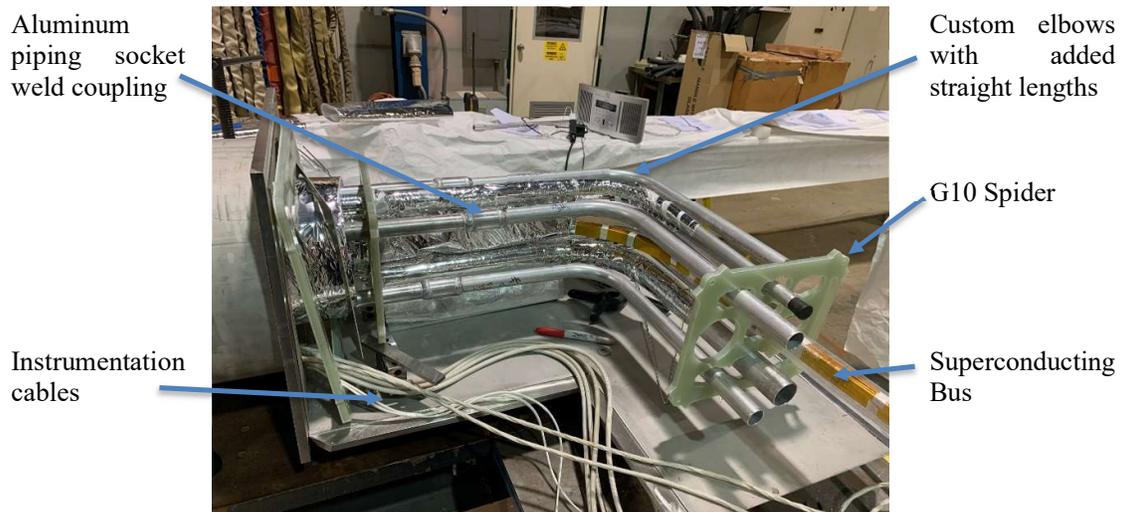

**Figure 10.** Photo showing STL right-angle joint internal piping, bus, supports, and instrumentation

The STL corner joint vacuum jackets were designed using flat plates due to space constraints. Code design calculations for rectangular flat plate vessels result in greater minimum wall thicknesses relative to cylindrical shells under external pressure. On the straight STL sections 12" Sch 10 pipe (324 mm OD x 4.57 mm W) was used. The flat plates used for the corner joints have a thickness of 12.7 mm. The total weld length for the flat plate design is also much greater than if space permitted a large pipe elbow. The GTAW procedures typically used by Fermilab welders on vacuum jackets resulted in very long welding times for each flat plate weld joint. New Gas Metal Arc Welding (GMAW) procedures and qualifications were created to reduce the welding time required for these flat plate weld joints. The GMAW process has significantly higher weld metal deposition rates relative to the GTAW process. GTAW was still used for the root pass and GMAW was used for all additional weld passes.

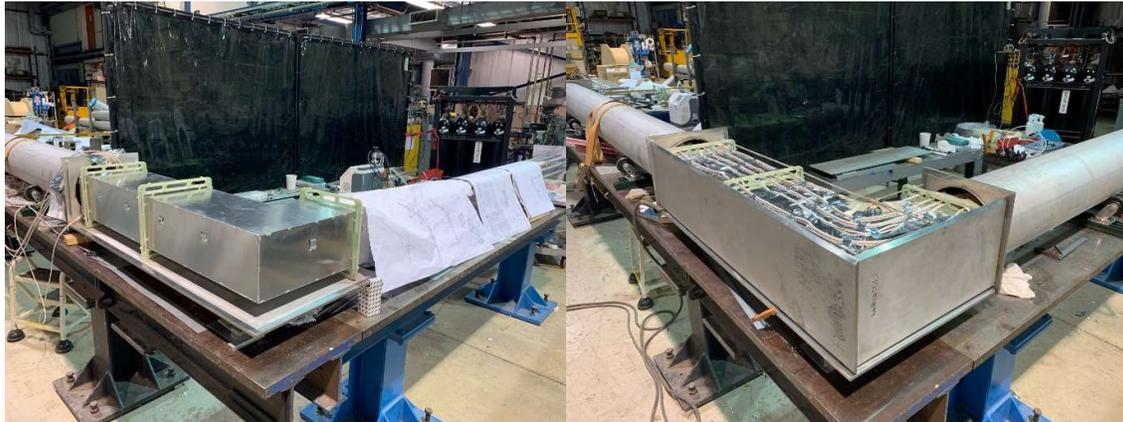

**Figure 11.** Photos showing installation of the STL corner joint thermal shield (left) and vacuum jacket (right)

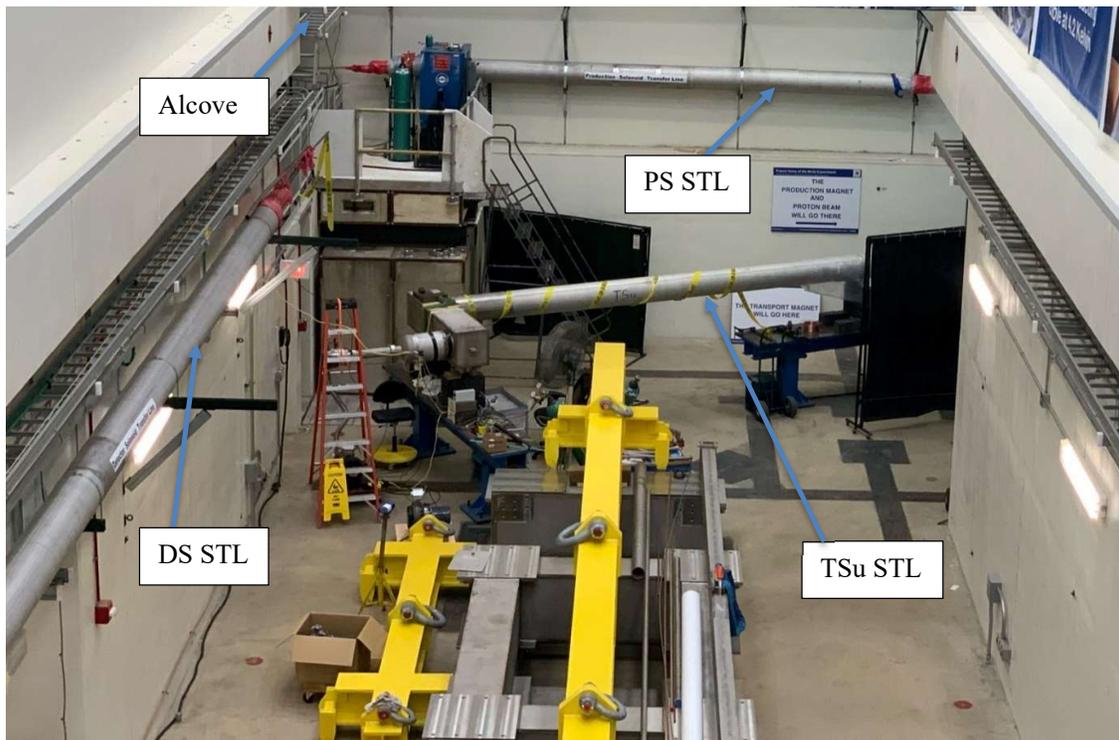

**Figure 12.** Photo showing the progress on STL installation in the lower detector hall of the Mu2e Building. The PS and DS horizontal STL sections have been mounted in their final locations. The TSu horizontal STL is undergoing a helium leak check in the photo above prior to being moved into its final location. The Feedboxes are located along the wall in the Solenoid Power Room above the alcove.

**Acknowledgments**
This manuscript has been authored by Fermi Research Alliance, LLC under Contract No. DE-AC02-07CH11359 with the U.S. Department of Energy, Office of Science, Office of High Energy Physics.